\documentstyle[twocolumn,aps,prl]{revtex}
\begin{document}
\title{Quantum Mechanics, Chaos and the Bohm Theory}
% \draft command makes pacs numbers print
\draft
% repeat the \author\address pair as needed
\author{Z. Malik}
\author{C. Dewdney}
\address{Department of Applied Physics, University of Portsmouth,
Portsmouth, PO1 2DZ, England}

\date{\today}
\maketitle
\begin{abstract}
% insert abstract here
The quantum counterpart of the classically chaotic kicked rotor is
investigated using Bohm's appraoch to quantum theory
\end{abstract}
% insert suggested PACS numbers in braces on next line

% body of paper here

\section{Introduction}

Formally Bohm's Causal Interpretation~\cite{bohm52}, or de Broglie's
Pilot wave~\cite{debroglie27} interpretation, of quantum mechanics
arises when the substitution $\psi= Re^{iS/\hbar}$ is made in the
Schr\" {o}dinger equation:
\begin{equation}
 \left
(-\frac{\hbar^{2}}{2m}\nabla^2 + V \right)\psi
=i\hbar\frac{\partial\psi}{\partial t} \label{eq:schro}
\end{equation}
and the real and imaginary parts are separated yielding the equations,
\begin{equation}
 \frac{\partial\rho}{\partial t} +\nabla\cdot(\rho{\bf v}) =0
\label{eq:con}
\end{equation}
and
\begin{equation}
-\frac{\partial S}{\partial t} = -
\frac{\hbar^{2}}{2m}\frac{\nabla^{2}R}{R} +
\frac{(\nabla S)^{2}}{2m} + V. \label{eq:hj}
\end{equation}
Where $\rho=\vert\psi\vert^{2} $ and
\begin{equation}
 m{\bf v}=\nabla S. \label{eq:vel}
\end{equation}
The particle is
assumed to have a definite, but unknown, position with a momentum
given by (\ref{eq:vel}).  Equation~(\ref{eq:hj}) can be interpreted as
a Hamilton-Jacobi-type of equation with an extra `quantum potential'
term Q, $Q= -\frac{\hbar^{2}} {2m} \frac{\nabla^{2}R}{R}$.  Newton's
second law is modified by the quantum force term and becomes
\begin{equation}
{\bf F} = -\nabla(Q+V) = -\nabla V_{\mbox{eff}}.
\label{eq:for}
 \end{equation}
The particle trajectories are the integral curves
of~(\ref{eq:vel}) and since S is determined by~(\ref{eq:schro}), both
the trajectory of an individual particle and the evolution of its
dynamical variables are determined by the time development of
$\psi$. Many detailed calculations have now been carried out
demonstrating exactly how the Bohm approach works in specific
cases~\cite{dh82,dkv84,dewdney85,dhk86,dhk87,dewdney87,ld90} and some
of these are summarised in Holland \cite{book:holland} and in Bohm and
Hiley \cite{book:bh93}. These examples amply illustrate the fact that
all of the dynamical variables associated with a given system have
well- defined values and that they evolve continuously according to
deterministic equations of motion. As a result, in Bohm¢s approach
to quantum mechanics, one can have recourse to all of the usual
apparatus of classical mechanics in characterising dynamical
behaviour.

In practise to solve for the motion of a system one integrates
(\ref{eq:vel}), however in principle, just as in classical mechanics,
one can also solve for the motion by integrating the modified Newton's
equation (\ref{eq:for}).  However, there are vital differences with
the situation in Classical Mechanics. Firstly, there is the nature of
the quantum potential which is not a pre-assigned function of the
system coordinates and can only be derived from a knowledge of the
wavefunction.  (In general even for static V's the quantum potential
will be both time- dependent and highly complex in form.) Secondly,
even when the quantum potential for a given system is known, one has
to impose the initial constraint that $p_{0}=\nabla S_{0}$ in the
solution of (\ref{eq:for}) to recover the quantum trajectories
calculated from (\ref{eq:vel}).

Looking at equation (\ref{eq:for}), and considering the highly
non-linear character of Q, one might, at first sight, expect chaotic
behaviour to arise.  Indeed some authors have suggested that the
trajectories may be expected to be chaotic or very sensitive to
initial conditions~\cite{book:bh93} but this seems to be at odds with
the well- known quantum suppression of classical chaos. In this paper
we hope to clarify this situation by investigating the quantum
counterpart of a paradigmatic example of classical chaos (the kicked
rotor) from Bohm's point of view. The indicator of chaos that we will
apply is the exponential divergence of trajectories, accompanied by
global confinement, in phase space. Evidently in the conventional
interpretation of quantum mechanics such an approach cannot be taken.

\section{The kicked rotor}

The kicked rotor has been extensively studied~\cite{ccif79,izrailev90}
in a variety of forms, but here we take as the hamiltonian
\begin{equation}
 H=p^2_{\theta}/2I-I\omega^2_0\cos{\theta}\sum_{n=-
\infty}^{\infty}\delta(t/T-n), \label{kikham}
\end{equation}
where
$I=ml^2$ is the inertia ($m$, $l$ being the mass and length of the
rotor), $\theta$ is the orientation, $p_{\theta}$ is the angular
momentum, $\omega_0=\sqrt{g/l}$ is the natural frequency for small
displacements, and $T$ is the period of the delta-function kicking.

Classically this hamiltonian describes a rotor (eg. a pendulum)
subjected to a periodically pulsed gravitational field, and it leads
to equations of motion which can be reduced to the Standard
Map~\cite{chirikov79}, (which exhibits all the features of a chaotic
system). The dynamics of the rotor depend on just one parameter
$K=(\omega_0T)^2$. For $K<<1$ the behaviour is regular and for $K>>1$
the motion is chaotic.

Quantum mechanically, an arbitray rotor wavefunction can be expanded
in terms of the free rotor simultaneous eigenfunctions of energy and
angular momentum according to
\begin{equation}
\psi(\theta,t)=\frac{1}{\sqrt{2\pi}}\sum_{n=-
\infty}^{\infty}a_n(t)e^{in\theta}.  \label{sup}
\end{equation}
The
time evolution of the wavefunction is then reduced to an iteration of
the expansion coefficients according to,
\begin{equation}
a_n((N+1)T^+)=\sum_{r=-\infty}^{\infty}a_r(NT^+)i^{n-r}J_{n-
r}(k)e^{-ir^2\tau/2}, \label{anp1exp}
\end{equation}
where $\tau=\hbar
T/I$, $k=(I\omega_0^2T/\hbar)$ and $J_s(k)$ are ordinary bessel
functions of the first kind. Equation~(\ref{anp1exp}) gives the value
of the expansion coefficients just after the $(N+1)^{th}$ kick in
terms of the coefficients just after the $N^{th}$ kick. In practice
the summation can be truncated, the bessel functions $J_s(k)$ become
negligible outside the range $s\approx 2k$ and we can check the
accuracy of the truncation by monitoring the conservation of the
probability.

Classically, in the chaotic regime (for, say $K=5$), the average
energy grows linearly with time showing a random diffusive-like
behaviour of the system, a possible indicator of chaos.  Quantum
mechanically, however, this growth of $\langle
E\rangle=\frac{\hbar^2}{2I}\sum_{n}n^2\vert a_n\vert^2$ is suppressed
after a certain time~\cite{ccif79}. This suppression occurs in the
semiclassical regime where $k=10$, $\tau=1/2$ and $K=5$.

\section{The Free Rotor and the Bohm Theory}

Between kicks the rotor is free and so we begin with an investigation
of the Bohm dynamics of the free rotor in various states and then
continue with the kicked case.  For a simultaneous eigenstate of
energy and angular momentum of the free rotor,
\begin{equation}
\psi(\theta,t)=\frac{1}{\sqrt{2\pi}}e^{-i\hbar
n_0^2t/2I}e^{in_0\theta},
\end{equation}
and the Bohm motion is a
subset of the classical motion, the angular momemtum of the rotor
using~(\ref{eq:vel}), is $p_{\theta}(t)=\hbar n_0$.

Next consider a simple superposition of the ground plus equal amounts
of the angular momentum $+1$ and $-1$ first excited states given by,
\begin{equation}
 \psi(\theta,t)=1+2a\cos{\theta}e^{-it/2},
\label{bpsifree4}
\end{equation}
where $a_1=a_{-1}=a$, $a$ is real and
set $\hbar=I=1$. The Bohm momentum for this state is readily shown to
be
\begin{equation}
p_{\theta}(t)=\frac{d\theta}{dt}=\frac{2a\sin{\theta}\sin{t/2}}
{1+4a^2\cos^{2}{\theta} +4a\cos{\theta}\cos{t/2}} \label{bp4}
\end{equation}
This can be integrated by making the substitution
$C=\cos{t/2}$, giving the following implicit trajectory equation
\begin{eqnarray} 4a\cos{t/2}(\sin{\theta_t}-\sin{\theta_0}) & = &
-(1+2a^2)(\theta_t-\theta_0) \nonumber \\
    & & \mbox{} - a^2(\sin{2\theta_t}- \sin{2\theta_0})
\label{bthetat4} \end{eqnarray} Equation~(\ref{bthetat4}) can be seen
to be periodic in time, with a period $T=4\pi$. Hence here, obviously
there can be no divergence in the trajectories.

An alternative approach to the calculation of the system's motion is
to attempt the direct numerical solution of
\begin{equation}
F=I\frac{\partial^2\theta}{\partial t^2}=-\frac{\partial
Q}{\partial\theta}.  \label{forceq}
\end{equation}
with Q derived from
(\ref{bpsifree4}), for this simple superposition state. If the initial
condition $p_{\theta}=\nabla S$ is imposed one recovers the
trajectories obtained above (\ref{bthetat4}). Without this initial
condition the trajectories appear to diverge
rapidly. Figure~\ref{fig1} shows the divergent behaviour of two
initially close trajectories in this case. Classically, for a given
potential and given initial conditions a trajectory may wander over
the whole of phase space whereas quantum mechanically it is restricted
to $p=\nabla S$~\cite{book:holland}. Trajectories with initial
conditions satisfying the latter condition, because of the
single-valuedness of the phase, never cross in configuration
space-time, but if this constraint is relaxed this feature is lost
allowing for the possibility of divergence.  Bohm trajectories for
superposition states can be exceedingly complicated, but for an
arbitrary superpostion we would expect the Bohm trajectories to be
quasiperiodic. Calculation shows no discernible divergence of
trajectories for such a superposition state with arbitrarily chosen
$a_n$'s.

\section{The kicked rotor and the Bohm theory}

We also calculated the trajectories for the kicked rotor, with
particular interest in the semiclassical regime, where $k=10$ and
$\tau=1/2$. The intial state of the rotor was taken to be the ground
state (giving an initial Bohm momentum of zero) and the trajectories
for two initially close positions were calculated.(Figures ~\ref{fig2}
and ~\ref{fig3}are poincare sections with a period equal to that of
the kick.)  Figure ~\ref{fig2} shows that there is practically no
divergence between two trajectories started at $\theta=30^{\circ}$ and
$\theta=30.5^{\circ}$ after eighty or so kicks. Figure~\ref{fig3}
shows two trajectories with the same initial position but with
slightly different wavefunctions (and therefore momenta): the two
trajectories are highly correlated. It turns out that even
trajectories which are far apart in orientation do not
diverge. Figure~\ref{fig4} shows the evolution of two trajectories
initially at $\theta=1^{\circ}$ and $\theta=30^{\circ}$, with the
initial state (for both) being a gaussian wavepacket.

Moreover the results above are typical for Bohm trajectories of the
kicked rotor system. Divergence in the trajectories cannot be obtained
by varying any of the parameters involved, including the choice of
initially highly excited states.  The trajectories themselves can be
exceedingly complicated, but there is no divergence.

Similar results follow for the Bohm trajectories with the
quasiperiodically kicked rotor~\cite{shepelyansky83,mag87,cis88}.

\section{Measurement}

We conclude by discussing the process of measurement in the Bohm
theory and its possible significance in a discussion of quantum
chaos. Chaotic systems are those
with only a few degrees of freedom in which practically random
behaviour arises, but not as a result of an external random
influence. In the usual interpretation of quantum mechanics
measurement is considered to be an inherently random external process
and because of this there is no detailed discussion of measured
systems in the quantum chaos literature. If one adopts Bohm's point of
view the situation is rather different. The Bohm theory has a
perfectly deterministic description of measurement as a dynamical
process and unique measurement results follow from unique initial
conditions.

The details of the Bohmian dynamics for an angular momentum
measurement using a Stern-Gerlach device have already been given, for
hydrogen-like atoms, in ~\cite{dm93}. Detailed work is now in progress
to investigate repeated angular momentum measurements on the kicked
rotor and on kicked spin systems, but some general observations are
relevant here. Recently D\"{u}rr et al~\cite{dgz92} have discussed
similar ideas, in the context of a somewhat different system.

We take the initial wave function of the rotor to be
\begin{equation}
\Psi(z,\theta,0)=\phi_0(z) \psi_0(\theta) \label{initprod}
\end{equation}
where $\phi(z)$ describes the position of the rotor and has the form
of a packet centred at $z=0$ and $\psi_0(\theta)$ is the ground state
of the oscillator. The Stern-Gerlach magnetic field gradients are
aligned with the z-direction and so only motion of the rotor in this
direction is significant

Consider now a sequence in which the system is repeatedly kicked and
then measured. Each time the system is kicked the wavefunction becomes
a superposition, ~(\ref{sup}), and each time a measurement is carried
out, providing we allow enough time for the wave packets associated
with different angular moment to separate along the z-axis, the
rotor's wave function will become effectively the simple product
\begin{equation}
 \Psi(z,\theta,t)=\phi_n(z,t)\psi_n(\theta,t)
\end{equation}
as the rotor's actual $z$-coordinate evolves to
uniquely associate it with just one of the $\phi_n$.  When this has
taken place all of the other terms in the superposition can be ignored
in the further stages of the calculation. (This is the equivalent of
the reduction of the wavepacket, the ``empty" packets do not affect
the motion of the system.)

At each measurement there will be a series of bifurcation points in
the inital packet at the entrance to the Stern-Gerlach apparatus, and
the actual value of $z$ with respect to these points will determine
which of the angular momentum eigenstates the system enters,
furthermore, the evolution of the wave-function under the kick (the
set of $a_n$ produced) depends on the wavefunction just before the
kick. Hence the sequence of eigenfunctions of angular momentum
produced in a particular sequence of kicks and measurements depends on
the precise value of the centre of mass coordinate ($z$) at the
outset.  The deterministic evolution of the rotor wave function
throughout the sequence of kicks and measurements can be calculated
and, since the particular evolution depends on the initial $z$-value,
the possibility exists for the evolution of the rotor wave functions,
for two slightly different initial z-values, to become divergent. If
the rotor wavefunctions for different initial conditions can become
divergent then the trajectories can cross (they obey $p=\nabla S$, but
$S$ devlops differently in each of the two cases) and we can look for
the signatures of chaos in the motion of the centre of mass or in the
rotation of the rotor.

In conclusion, it is clear that the Schr\" {o}dinger evolution of the
wave function with its probabilistic interpretation, which removes the
possibility to discuss individual systems, will not allow quantum
chaos. Even in our repeatedly measured and kicked system the evolution
of the complete wavefunction (with all components of the superposition
maintained at each stage) will not allow for divergence in the sense
that two initially close wavefunctions will remain close.Supplementing
the Schr\" {o}dinger description with the Bohm trajectory description
allows an acces to the motion of individual systems and we have seen
that two isolated systems with slightly different initial conditions,
either initial position or initial wavefunction, will not diverge. The
possibility to produce divergent effective wave functions for
individual systems with slightly different initial conditions does
arise if we consider interactions with the system. An alternative to
measurement is to consider interaction with an environment
\cite{srpreprint}, but from Bohm's point of view.

% now the references. delete or change fake bibitem. delete next three
% lines and directly read in your .bbl file if you use bibtex.

\begin{figure}[h] \protect\caption{Momentum/time plot for the
superposition state $\psi(\theta,t)=1+2a\cos{\theta}e^{-it/2}$, of the
free rotor when the initial condition $p_{\theta}=\nabla\theta$ is
relaxed. Initially $\theta=1^{\circ}$ or $\theta=1.01^{\circ}$ and
$\dot{\theta}=1$.}  \protect\label{fig1} \end{figure}

\begin{figure}[h] \protect\caption{Orientation/time plot for the
kicked rotor with $\omega_0=(20)^{\frac{1}{2}}$ and $T=1/2$. The two
trajectories are initially close together with initial angles
$\theta=30^{\circ}$ and $\theta=30.5^{\circ}$.  The rotor is initially
in the ground state $a_0=1$.}  \protect\label{fig2} \end{figure}

\begin{figure}[h] \protect\caption{ Orientation/time plot for the
kicked rotor with $\omega_0=(20)^{\frac{1}{2}}$ and $T=1/2$. Here we
have two trajectories given by the initial states,
$a_0=1,a_1=(4637/13313)^{\frac{1}{2}}$ and
$a_0=1,a_1=(0.3483)^{\frac{1}{2}}$ and both with the same initial
angle, $\theta=30^{\circ}$.}  \protect\label{fig3} \end{figure}

\begin{figure}[h] \protect\caption{ Orientation/time plot for the
kicked rotor with $\omega_0=(20)^{\frac{1}{2}}$ and $T=1/2$. Two
trajectories are taken initially {\em far apart} with initial angles
$\theta=1^{\circ}$ and $\theta=30^{\circ}$. The initial state is a
gaussian wavepacket with a momentum half-width 0.5 and an initial
momentum of two.}  \protect\label{fig4} \end{figure}

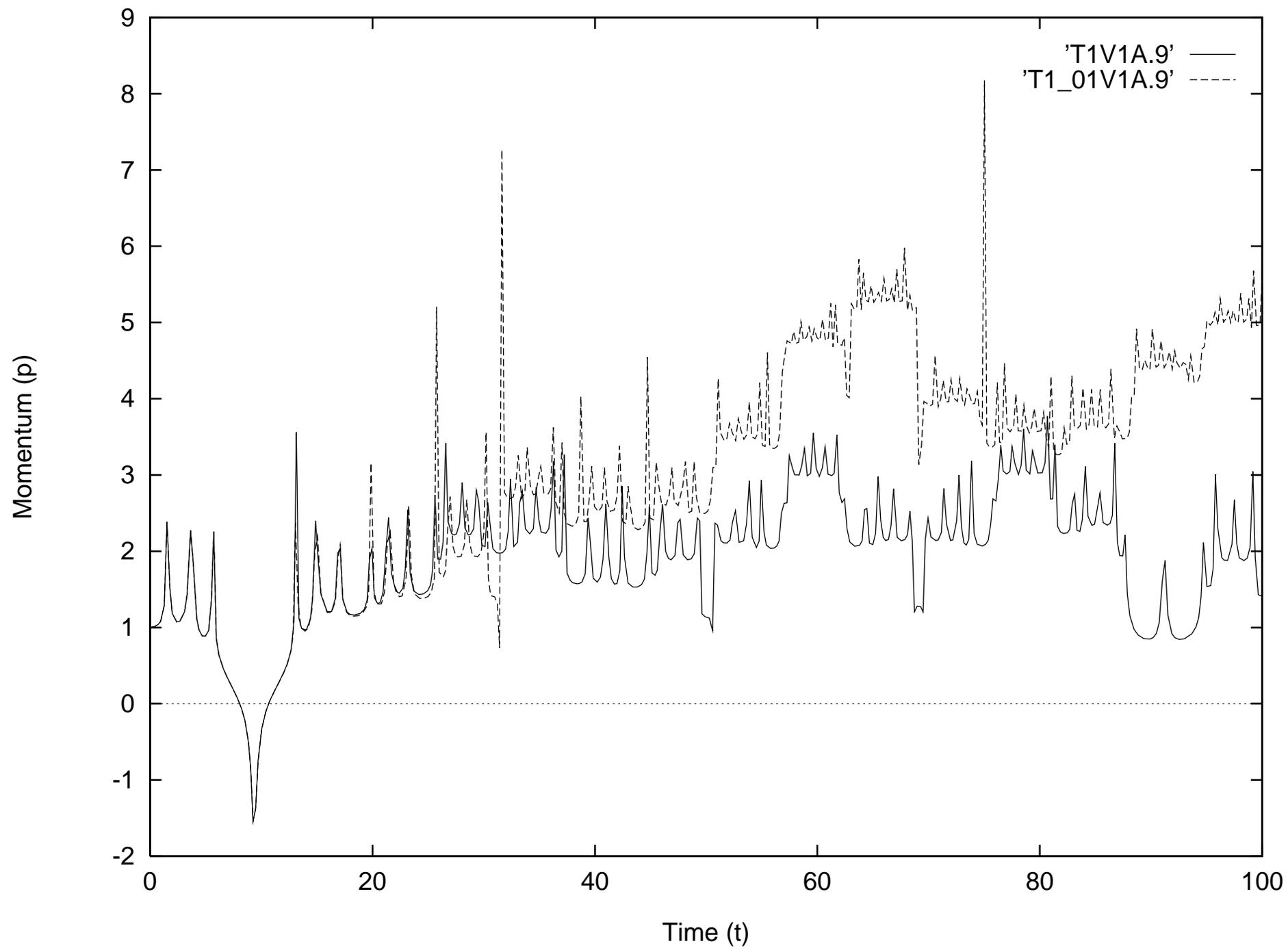
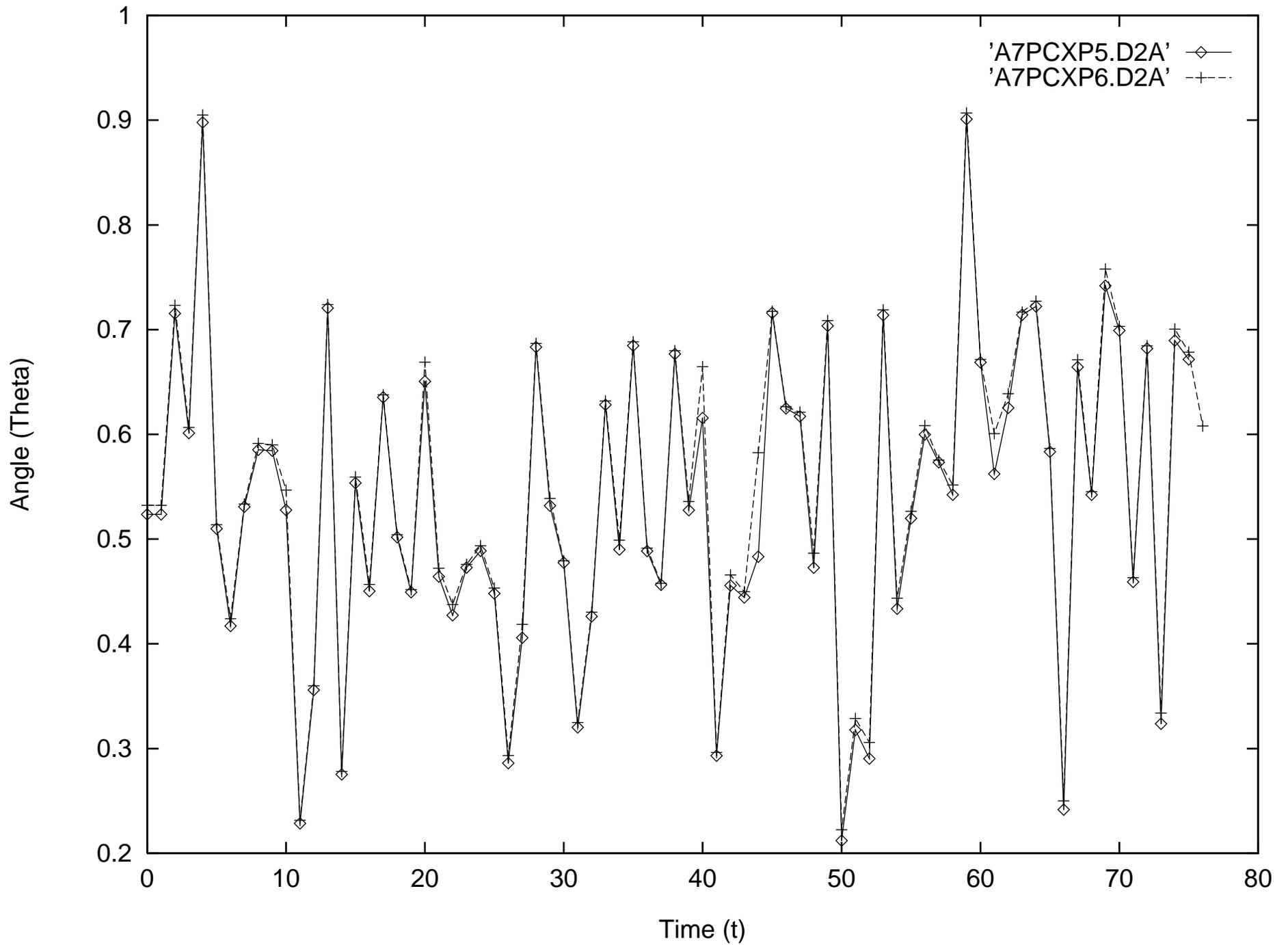
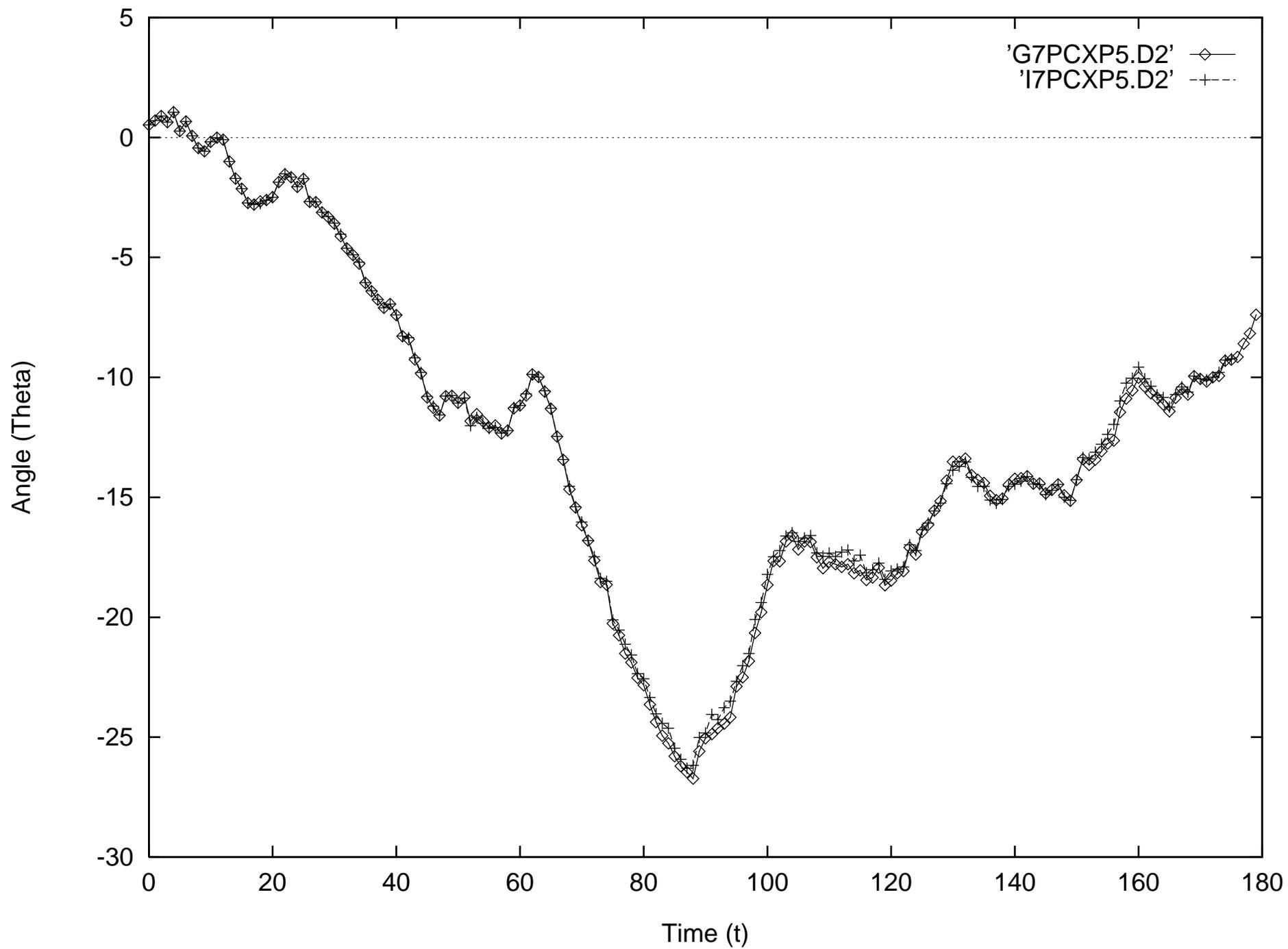
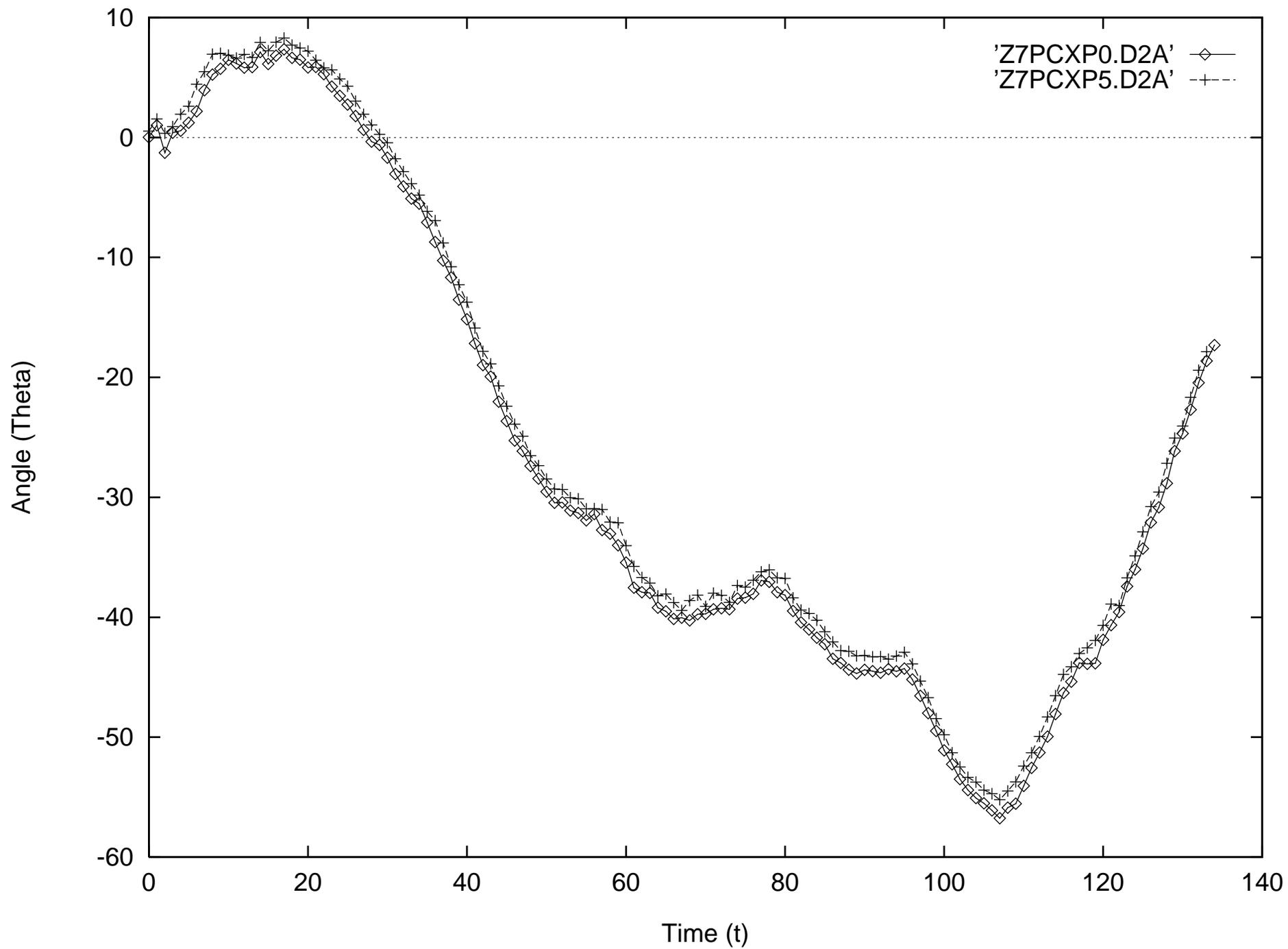
\end{document}